\documentclass{aastex}
\usepackage{emulateapj5}

\newcommand{\pmnj}{PMN~J0134$-$0931}
\newcommand{\jay}{J0134-0931}

\newcommand{\civ}{C\,{\sc iv}}

\newcommand{\kms}{km\,s$^{-1}$}

\newcommand{\mgii}{Mg\,{\sc ii}}

\newcommand{\oii}{[O\,{\sc ii}]}
\newcommand{\oiii}{[O\,{\sc iii}]}

\newcommand{\ciii}{C\,{\sc iii}]}

\newcommand{\caii}{Ca\,{\sc ii}}

\newcommand{\feii}{Fe\,{\sc ii}}

\begin{document}

\journalinfo{
Accepted to ApJL on July 11, 2002} 
\submitted{}
\title{The Redshift of a Lensing Galaxy in PMN~J0134$-$0931}

\author{
Patrick B. Hall,\altaffilmark{1,2}
Gordon T. Richards,\altaffilmark{3}
Donald G. York,\altaffilmark{4,5}
Charles R. Keeton,\altaffilmark{4}
David V. Bowen,\altaffilmark{1}
Donald P. Schneider,\altaffilmark{3}
David J. Schlegel,\altaffilmark{1}
J. Brinkmann\altaffilmark{6}
}
\altaffiltext{1}{Princeton University Observatory, Princeton, NJ 08544-1001}
\altaffiltext{2}{Departamento de Astronom\'{\i}a y Astrof\'{\i}sica,
Facultad de F\'{\i}sica, Pontificia Universidad Cat\'{o}lica de Chile, 
Casilla 306, Santiago 22, Chile}
\altaffiltext{3}{Department of Astronomy and Astrophysics,
The Pennsylvania State University, University Park, PA 16802}
\altaffiltext{4}{Department of Astronomy and Astrophysics, 
The University of Chicago, 5640 S. Ellis Ave., Chicago, IL 60637}
\altaffiltext{5}{Enrico Fermi Institute, The University of Chicago, 
5640 S. Ellis Ave., Chicago, IL 60637}
\altaffiltext{6}{Apache Point Observatory, P.O. Box 59, Sunspot, NM 88349-0059}

\begin{abstract}

The Sloan Digital Sky Survey (SDSS) automatically targeted as a quasar
candidate the recently discovered, gravitationally lensed, extremely reddened
$z=2.2$ quasar \pmnj.
The SDSS spectrum exhibits \caii\ absorption at $z=0.76451$,
which we identify as the redshift of a lensing galaxy.
{\em Hubble Space Telescope} imaging shows that components CDE of the system are
significantly redder than components A or B and detects faint galaxy emission 
between D and A+B.
The redshift of the dust responsible for the reddening 
remains unconstrained with current data. 
However, we outline a model wherein lensing and differential reddening 
by a $z=0.76451$ galaxy pair can entirely explain this system.

\end{abstract}

\section{Introduction}  \label{INTRO}

The Sloan Digital Sky Survey\footnote{The SDSS Web site is http://www.sdss.org/.}
\markcite{yor00}(SDSS; {York} {et~al.} 2000) is using a drift-scanning imaging camera 
\markcite{gun98}({Gunn} {et~al.} 1998) to image 10$^4$\,deg$^2$ of sky on the SDSS
$ugriz$ AB magnitude system \markcite{fuk96,sdss82,sdss105}({Fukugita} {et~al.} 1996; {Hogg} {et~al.} 2001; {Smith} {et~al.} 2002).
Two multi-fiber, double spectrographs on a dedicated 2.5m telescope
are used to obtain spectra for
$\sim$10$^6$ galaxies to $r=17.8$ and $\sim$10$^5$ quasars to $i=19.1$
($i=20.2$ for $z>3$ candidates).  As discussed in \markcite{sdssqtarget}{Richards} {et~al.} (2002),
quasar candidates are targeted for spectroscopy
because they are outliers from the stellar locus
or because they are unresolved objects
with radio emission detected by the FIRST survey \markcite{bwh95}({Becker}, {White}, \& {Helfand} 1995).
Due to these inclusive criteria and its area and depth, the SDSS is effective
at finding quasars with unusual properties and colors (\markcite{sdss123}{Hall} {et~al.} 2002). 

Quasars heavily reddened by dust form one population of quasars with `unusual'
colors, at least for optically selected, magnitude limited samples.  Recently,
\markcite{gre01}{Gregg} {et~al.} (2002), hereafter G02, reported on two heavily reddened quasars from a
survey of FIRST radio sources with red counterparts in the Two-Micron All-Sky
Survey \markcite{skr97}({Skrutskie} {et~al.} 1997).  One of these --- \pmnj\ --- is gravitationally lensed,
as discovered independently by G02 and \markcite{win02}{Winn} {et~al.} (2002), hereafter W02.

W02 identify six radio components of \jay, five of which have the same spectral
slope (not measurable for the faintest component, F),
and two of which (C and E) have lower surface brightnesses than the others.
W02 present two lens models.  In one, \jay\ is a six-image lens,
with C and E differentially broadened by interstellar scattering 
(\markcite{jon96}{Jones} {et~al.} 1996).
This model requires more than one lensing galaxy and even then might not 
reproduce the image configuration.  In the second,
despite the similar spectral slopes of all the radio components,
C and E are a foreground object or objects
while A+B and D+F are the double images of a core+jet source.
This model requires almost perfect source-lens alignment
to match the flux ratios.  However, 
components A, B and D+F (the D+F separation is only 0\farcs05)
have near-IR (NIR) counterparts with very different flux ratios
than in the radio.  More data are clearly needed for a viable lens model.

\jay\ was independently targeted as a quasar candidate in the SDSS.
Here we investigate this system further using SDSS
and {\em Hubble Space Telescope} ({\em HST}) data.

\section{SDSS Data on \pmnj}  \label{SDSS} 

\jay\ was bright enough in the SDSS for consideration as a high-redshift
quasar candidate (Table \ref{t_info}).  It qualifies as such a candidate
because it has $g^*-r^*=2.26$, redder than the $g^*-r^*>2.1$ threshold used
as part of the $z\geq3.6$ quasar candidate selection \markcite{sdssqtarget}({Richards} {et~al.} 2002).

\begin{figure}
\epsscale{1.00}
\plotone{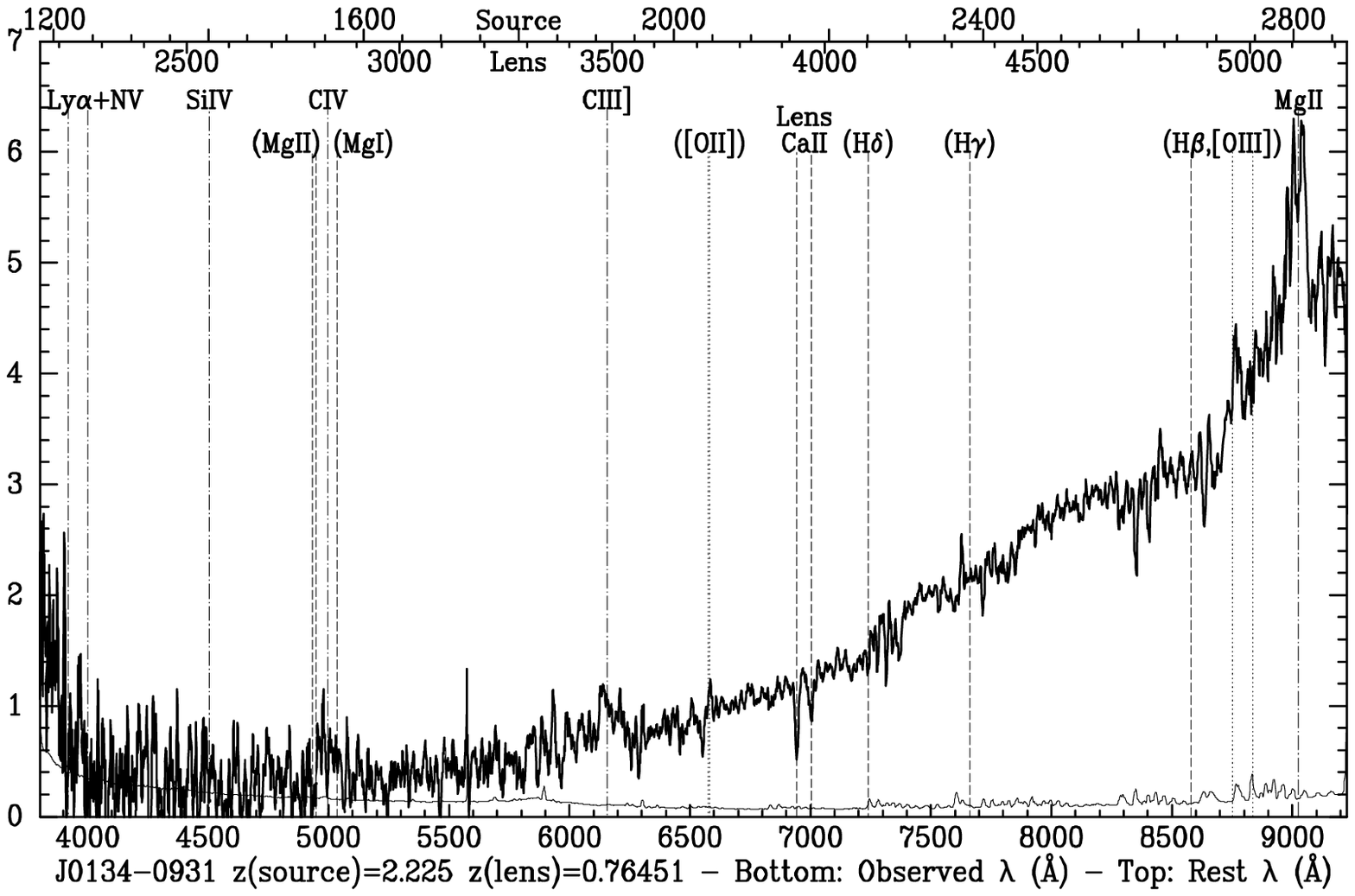}
\caption[]{
\singlespace 
The coadded SDSS spectrum ($\lambda/\delta\lambda$$\simeq$1800) of \jay,
smoothed by a seven pixel boxcar.  The noise level is shown by
the thin line across the bottom of the Figure.
The abscissa is $F_{\lambda}$ in units of
$10^{-17}$\,ergs cm$^{-2}$\,s$^{-1}$\,\AA$^{-1}$.
Observed wavelengths (in \AA) are shown below
the bottom axis, and wavelengths in the rest frame of the 
source (lens) above (below) the top axis.  Dash-dot lines show
the expected wavelengths of broad emission lines at $z=2.225$.
Dashed lines show \caii\ absorption at $z=0.7645$ as well as
the expected wavelengths of strong lines from H and Mg at $z=0.76451$.
Dotted lines show
the expected wavelengths of narrow \oii\ and \oiii\ emission at $z=0.76451$.
}\label{f_spec}
\end{figure}

The SDSS obtained spectra of \jay\ on UT Aug. 26 and Sep. 26, 2001, for
a total exposure of 125.1 minutes.
The inverse-variance-weighted average spectrum,
smoothed by a seven pixel boxcar, is shown in Figure\,\ref{f_spec}.
Broad \mgii\ emission from the quasar is seen, along with weak \ciii\ and \civ.
The $\lambda$$\lesssim$3900\,\AA\ rise
is spurious: it is within the range of observed sky fiber residuals, and 
photometry shows this object is much fainter in $u$ ($<$3900\,\AA) 
than in $g$ (3900--5400\,\AA).  Cross-correlation
with the SDSS composite quasar of \markcite{sdss73}{Vanden Berk} {et~al.} (2001) yields a
redshift $z=2.225\pm0.006$, which is 840\,\kms\ higher than G02's $z=2.216$
based on the optical/NIR spectrum.

Also present in the SDSS spectrum are two absorption lines near 7000\,\AA.
Both lines are detected in each of the two SDSS spectra, and
can in fact be seen in Figure\,3 of G02.
The wavelengths match those of \caii\,H\&K at
$z=0.76451 \pm 0.00016$ and~do~not~match~intervening
\feii\ or \mgii.~The absorption is resolved,
with intrinsic $\sigma_v = 220 \pm 40$\,\kms,
and the EW ratio 0.55$\pm$0.16 is consistent with predominantly unsaturated
absorption~(EW ratio 0.5).\footnote{All uncertainties calculated from
resampling a two-gaussian fit to the data, constrained to one
FWHM and the \caii\ wavelength ratio, with IRAF {\sc ngaussfits}. 
The Image Reduction and Analysis Facility
is distributed by NOAO, operated by AURA, Inc., under contract to the NSF.}
The absorption is very strong, with rest-frame
$EW_{\rm CaII,K}=5.3\pm0.5$\,\AA, vs. typical
$EW$$<$1\,\AA\ seen in the Galaxy and extragalactic \mgii\ absorption systems 
(\S\,IV~of \markcite{bwts85}{Briggs} {et~al.} 1985). The only objects with similar values 
are\,the\,lens\,B\,0218+357\,\markcite{bpww93}({Browne} {et~al.} 1993)\,and\,the\,ultra-luminous~IR~galaxies~of~\markcite{rvs02}{Rupke}, {Veilleux}, \& {Sanders} (2002).

The smoothed, coadded spectrum shows
{\em apparent} absorption and emission, respectively, near the expected
positions of \mgii\ and \oii\ at $z=0.7645$.
However, inspection of both unsmoothed spectra
shows that neither feature was obviously present on both spectra, 
and that other apparent narrow emission and absorption features
(e.g., at 8300--8900\,\AA)
are sky line subtraction residuals or lie near sky lines, 
where the noise is higher.
Thus we do not believe there is any significant, narrow absorption or emission 
in the SDSS spectrum, except for \caii.

\section{HST Data on \pmnj}  \label{HST} 

\begin{figure}
\epsscale{1.05}
\plotone{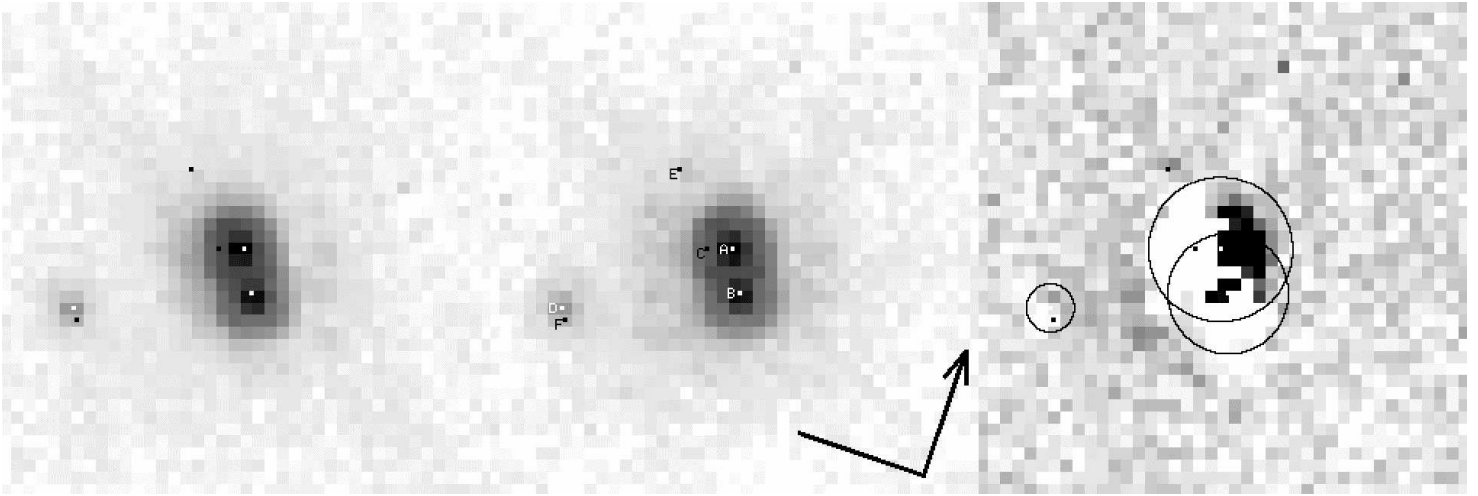}
\caption[]{
\singlespace 
The center panel shows the
{\em HST} {\em F814W} planetary camera image of \jay\
(0\farcs046/pixel). 
North (with arrow) and East are shown with bars 0\farcs50 long.
The six radio components are labelled: components whose
positions were measured in the {\em HST} image are white,
and those whose positions were inferred from radio data are black.
The left panel shows a PSF image containing only the emission expected
from point sources at A, B and D, displayed on the same
log(square~root) intensity scale as the raw image.
All panels are reverse contrast (dark areas are emission).
The right panel shows the residuals after subtraction of the PSF image,
displayed on a linear intensity scale. 
The circles outline regions dominated by PSF subtraction residuals
The integer pixel shifts used to make the PSF image 
inadequately account for the subpixel centroids of A, B and D.
Nonetheless, beyond the extent of these
residuals, there is faint but real emission between D and A+B.
}\label{f_img}
\end{figure}

{\em HST} WFPC2 images of \jay\ in {\em F814W} and {\em F555W} were retrieved
from MAST.\footnote{G.O. \#9133, P.I. Falco. MAST is the Multimission Archive
at STScI, which is operated by AURA, Inc., under NASA contract NAS5-26555.}
In each filter, two undithered 21.7 minute exposures were coadded
with iterative cosmic ray rejection.  Only components ABD
are clearly detected in {\em F814W} (Figure\,\ref{f_img}, center panel),
and only A and B in {\em F555W} (not shown).  Components CE are not present
at the flux ratios relative to A and B seen in the NIR or radio,
and component D is also unexpectedly faint.
Using IRAF {\sc qphot} and 0\farcs092 (2 pixel) radius apertures,
we measure A:B:D flux ratios accurate to $\sim$5\%
of 0.93:1:$<$0.025 in $V_{555W}$ and 1.45:1:0.044 in $I_{814W}$,
vs. 1.11:1:0.28 in $K'$ and 4.05:1:0.76 at 3.6\,cm (see G02).

To determine how much of the low-surface-brightness emission around \jay\ is
due to the point-spread functions (PSFs) of A+B, we used two unsaturated stars
on the planetary camera.  Each star was rescaled to the same flux as component
A, B or D in a 0\farcs092 radius.  These three scaled images
were then shifted by integer pixels to the positions
of the corresponding components and coadded.
The averaged PSF image of the two stars is shown in the left panel of
Figure\,\ref{f_img}, and the PSF-subtracted image of \jay\ 
in the right panel.  The circles show areas
with significant PSF subtraction residuals.
We see possible emission at the position of component E, and a definite
excess of low-surface-brightness emission between components D and A+B.
Accurately determining the centroid and morphology of the extended emission
will require a subpixel shift analysis of the {\em HST} images,
since our PSF subtraction residuals prevent detection of emission
within $\lesssim$0\farcs25 of A+B.  (Note that G02 did find possible emission
midway between components A and D in their deconvolved $K'$ image.)
The total extended flux we can detect is $\sim$4.1\% of the 0\farcs092 radius
$I_{814W}$ flux of component B.

\section{Where is the Dust?}  \label{DUST} 

We estimate the redshift and amount of dust reddening in \jay\ by reddening
a composite quasar constructed from the SDSS Early Data Release quasar
catalog \markcite{sdssedrq}({Schneider} {et~al.} 2002), following \markcite{sdss73}{Vanden Berk} {et~al.} (2001).  We minimized
the $\chi^2$ between the reddened composite
and the unsmoothed SDSS spectrum of \jay\ (after correcting the latter for 
Galactic extinction using IRAF {\sc deredden}), accounting for both
the observed errors on the fluxes in each pixel and the rms variation
of the composite quasar at the corresponding rest-frame wavelength.
We considered only the SMC reddening curve \markcite{pei92}({Pei} 1992), since G02 found
that it was the only empirical curve which could reasonably match
the spectrum of \jay.

If the dust is at the quasar redshift, we find
$E(B-V)=0.670\pm0.015$, with $\chi^2_{\nu}=1.073$.
With $R_V=2.93$ \markcite{pei92}({Pei} 1992), this corresponds to $A_V=1.96\pm0.05$,
in good agreement with G02 ($A_V=2.1$).
If the dust is at $z=0.7645$, we find 
$E(B-V)=1.315\pm0.025$ with $\chi^2_{\nu}=1.041$.
This corresponds to $A_V=3.85\pm0.08$, 
lower than the $A_V=4.5$ found by G02 when they assumed $z_{lens}=0.5$.

\begin{figure}
\epsscale{0.60}
\plotone{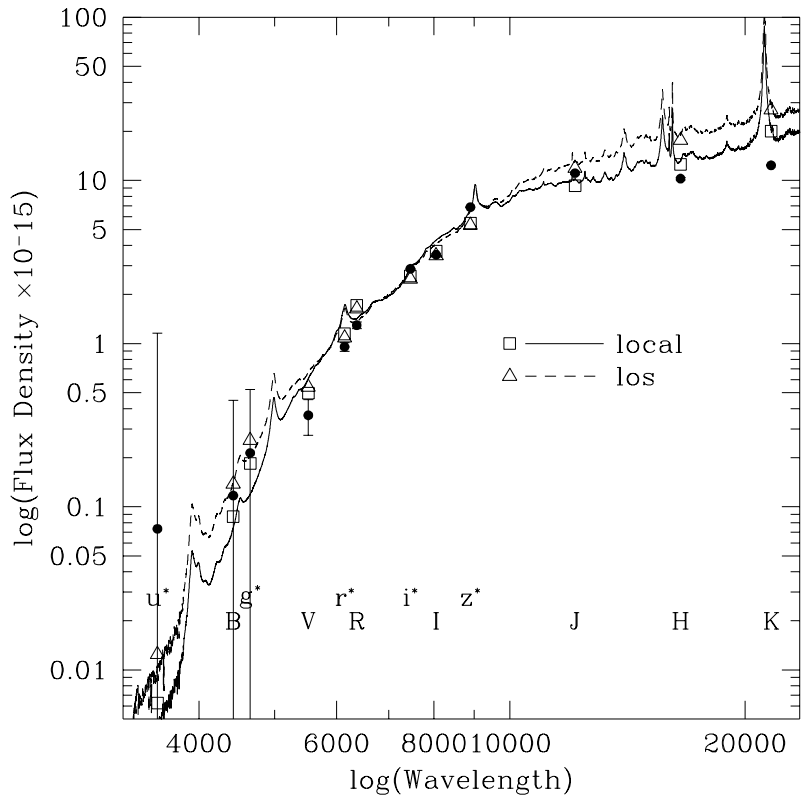}
\caption[]{
\singlespace 
`Best' fits of the SDSS EDR composite quasar reddened by 
$E(B-V)=0.67$ of dust at the quasar redshift (`local')
and by line-of-sight $E(B-V)=1.3$ at the lens redshift (`los')
to all published photometry of \jay\ (filled points with error bars).
(Since components A and B contribute essentially all the optical light but
only $\sim$84.5\% and $\sim$81.5\% of the $J$ and $K'$ emission (G02),
and by interpolation 83\% in $H$,
we reduced the 2MASS magnitudes accordingly for a consistent comparison.)
The normalizations have been adjusted to minimize the $\chi^2$ compared
to the optical/NIR photometry.  Neither fit is statistically acceptable, 
but dust at the lens redshift (which best fits the optical spectrum) is a
considerably worse fit to the NIR photometry than dust at the quasar redshift.
}\label{f_fits}
\end{figure}

Both fits have $\chi^2_{\nu} \simeq 1$, but the goodness of fit parameter
\markcite{nrf}({Press} {et~al.} 1992) shows that dust at $z=0.7645$ is formally a better fit.
We synthesized {\em ugrizBVRIJHK$_s$} 
photometry of the best-fit reddened composite quasars to see if
the fits to the optical spectrum matched the NIR photometry.
Figure\,\ref{f_fits} shows that they do not, and~that~dust at $z$=0.7645
gives a worse fit.
For SMC reddening, \jay\ is as bright in $J$ as expected from its optical
colors, but not as red in $J-K_s$.  Variability between the NIR and optical
imaging epochs cannot explain this, as quasars are bluer when they are brighter
\markcite{tv02}({Tr{\` e}vese} \& {Vagnetti} 2002).

\section{A Plausible Model for \pmnj}  \label{DIS} 

The SDSS has detected extremely strong \caii\ absorption at
$z=0.76451\pm0.00016$ 
in the spectrum of the reddened, gravitationally lensed $z=2.225$ quasar \pmnj.
We identify this as absorption from a gas-rich lensing galaxy.
Current data do not allow us~to
discriminate between dust reddening at the source or in the lens.  However,
{\em HST} imaging shows that component D is $\simeq$2\fm3 redder in
$I_{814W}-K'$ than the already heavily reddened components A and B
which dominate the~flux~in the SDSS spectrum.
Spectroscopy of component D alone is lacking, 
but the radio data of W02 are strong evidence that at least A, B and D
are images of a single quasar.~If~so, differential reddening in an intervening 
galaxy seems the most plausible explanation for the colors of components D and B
(and probably C and E).
{\em HST} has also detected faint emission between A+B and D.  If the
$z$=0.7645 galaxy differentially reddens component D, such $I_{814W}$-band
emission, redward of the redshifted 4000\,\AA\ break, is expected.


Note that the strong \caii\ absorption indicates a high column density of gas at
$z=0.7645$, but not necessarily of dust, because \caii\ is 
easily depleted \markcite{rea88}({Robertson} {et~al.} 1988).  
Thus, differential reddening at $z=0.7645$ is not {\em required}, though
it is of course possible for physically unrelated dust reddening and
\caii\ absorption to occur along the same sightlines at $z=0.7645$.
The maximum \caii\ absorption depth of $\sim$75\% of the continuum
indicates that largely unsaturated absorption is present in both
A and B, since they contribute to the optical light in a ratio 
$\sim$3:2.\footnote{Both A and B being images of the quasar
argues against W02's model wherein they are images of a core and jet,
as the optical source would have to be larger than the core+jet.
This model would be ruled out if detailed analysis of
the {\em HST} data showed no optical arc from A to B.}

One simple hypothesis is that \jay\ is lensed by two galaxies (see below),
at least one at $z=0.7645$, wherein
differential scatter broadening produces the larger radio sizes of C and E
while differential extinction makes components BCDE much redder than A.
(Dust at the source may still play a role,
but cannot produce significant differential reddening between components.)
Differential reddening is seen in two-thirds of lenses \markcite{fea99}({Falco} {et~al.} 1999),
and was suggested by G02 as a possible explanation for 
the red $J-K'$ of component D compared to A or B.
Note that the dust column densities in front of CDE
must be large enough to produce 
$\Delta E(B-V)\gtrsim0.27$ for $z_{dust}<2.225$
($\Delta E(B-V)\gtrsim0.56$ at $z=0.7645$),
which occurs in only $\sim$10\% of lens sightlines \markcite{fea99}({Falco} {et~al.} 1999).

Even though we and G02 have shown that SMC, LMC or MW reddening
is a poor fit to the photometry of \jay,
given the diversity of astrophysical extinction curves \markcite{sm79}({Savage} \& {Mathis} 1979), 
in principle simple differential extinction at $z=0.7645$ --- with an extinction
curve steeper than the SMC's at $\lambda\lesssim5000$\,\AA\ --- could still 
explain the relative fluxes of all components except for A. 
The A:B ratio is 4:1 in the radio and 1.1:1 at $K'$.
	Explaining this difference via microlensing amplification of the radio
	but not the optical source in A can be ruled out by the lack of
	variability in the total radio flux over almost 20 yr (Figure 2 of W02).
If this difference was due to differential extinction 
with the same extinction curve as the other components,
the observed optical A:B ratio would be $\ll$1, which is not the case.
One possible explanation for A is a range of reddenings across 
its sightline through an intervening galaxy.  If 75\% of A's flux
was reddened as much as component D, that reddened flux would dominate at $K'$
while the less reddened 25\% would dominate at shorter wavelengths.
Such reddening gradients are seen in some quasars
(\markcite{sdss123}{Hall} {et~al.} 2002, \S\,5.3), 
but the $\lesssim$0.1\,pc size scale over which this reddening gradient
must occur is much smaller than the minimum cloud diameter 
of $\sim$3\,pc inferred by \markcite{frye97}{Frye}, {Welch}, \& {Broadhurst} (1997) for the lens in PKS\,1830$-$211.


To explain the image configuration in the differential reddening model, 
we suggest that D+F is a {\em single} image, showing the true source morphology
of a core and weak jet.  and that the jet images corresponding to
the core images A and B are missing because
the source plane positions of the core and jet
yield different numbers of multiple images.  
A two-deflector version of the five-deflector system in Figure\,2 of
\markcite{kee02}{Keeton} (2002) can produce five bright images in roughly the \jay\ configuration,
but only two images of a nearby jet (for certain position angles).
This model is quite testable: it requires a faint component G near C
which is a parity-reversed image of F;
E and C+G to be broadened by interstellar scattering
(and to a lesser extent A, since its radio surface brightness is lower than B);
and two lens galaxies, one near E and one between A and D,
both locations at which G02 saw possible $K'$ emission.

As\,for other\,tests, high\,spatial resolution\,NIR\,images and 
spectra could yield colors and positions for all deflectors and components,
determine which are images of the quasar (via H$\alpha$), and measure
extinction curves \markcite{fea99}({Falco} {et~al.} 1999).
Absolute and differential reddenings at $z=0.7645$ can be constrained by
looking for molecular absorption in the radio components (\markcite{frye97}{Frye} {et~al.} 1997).
A high signal-to-noise ratio spectrum could constrain the dust location via
\ion{K}{1}, C$_{\rm 2}$, \ion{Li}{1} or CN absorption directly from dusty gas.

This work shows that the SDSS can contribute to the study of reddened quasars.
G02 state that `such objects will be entirely missed by standard ... optical
quasar surveys', yet the SDSS selected \jay, by its color,
as a quasar candidate.  Such reddened, low-$z$ quasars
have colors similar to normal high-$z$ quasars \markcite{sdss81}({Richards} {et~al.} 2001), and
thus SDSS selects them a magnitude deeper than unreddened low-$z$ quasars.
Also, \jay\ would have been selected as a low-$z$ quasar candidate via both
its colors and FIRST, had it been only 0\fm523 brighter in $i^*$.
However, unlensed
objects as red as \jay\ will be rare at the magnitude limits of the SDSS,
which probes only the top of the luminosity function for the reddest objects.  
Thus, since AGN with reddenings larger than that of \jay\ exist,
a full understanding of the prevalence 
and properties of reddened quasars will require multiwavelength observations.
Nonetheless, the large area of the SDSS will enable study of the 
reddening distribution to at least $E(B-V)\simeq0.65$.
In G02's terminology, the SDSS is not a `standard' optical quasar survey, and
is quite sensitive to reddened quasars (Richards et al. 2002, in prep.).

\acknowledgements

We thank the anonymous referee for useful comments.  The SDSS
is a joint project of the University of
Chicago, Fermilab, the Institute for Advanced Study, the Japan Participation
Group, The Johns Hopkins University, Los Alamos National Laboratory,
the Max-Planck-Institute for Astronomy,
the Max-Planck-Institute for Astrophysics,
New Mexico State University, Princeton University, 
the United States Naval Observatory, and the
University of Washington.  Apache Point Observatory, site of the SDSS
telescopes, is operated by the Astrophysical Research Consortium.
Funding for the SDSS
has been provided by the Alfred P. Sloan Foundation,
SDSS member institutions, the National Aeronautics and Space 
Administration, the National Science Foundation, the U.S. Department of 
Energy, the Japanse Monbukagakusho, and the Max Planck Society.
P. B. H. is supported by
FONDECYT grant 1010981, and D.\,P.\,S. \& G.\,T.\,R. by NSF grant 99-00703.


\begin{deluxetable}{lcccccccc}
\tablecaption{SDSS Data on PMN~J0134$-$0931\label{t_info}}
\tabletypesize{\scriptsize}
\tablewidth{520.00000pt}
\tablehead{
\colhead{J2000} & \colhead{Source} & \colhead{Lens} &
\multicolumn{5}{c}{SDSS Photometry on UT Sep. 27, 2000} &
\colhead{Galactic}\\[.2ex]
\colhead{Coordinates} & \colhead{Redshift} & \colhead{Redshift} &
\colhead{$u^*\pm\sigma_{u^*}$} &
\colhead{$g^*\pm\sigma_{g^*}$} &
\colhead{$r^*\pm\sigma_{r^*}$} &
\colhead{$i^*\pm\sigma_{i^*}$} &
\colhead{$z^*\pm\sigma_{z^*}$} &
\colhead{$E$$($$B$$-$$V$$)$}
}
\startdata
013435.66$-$093102.9 & 2.225$\pm$0.006 & 0.76451$\pm$0.00016 & 
25.34$\pm$1.08 & 23.54$\pm$0.31 & 21.28$\pm$0.06 & 19.65$\pm$0.03 & 18.30$\pm$0.03 & 0.0305 \\
\enddata
\tablecomments{
The SDSS coordinates should be good to $\sim$0\farcs060 rms each for an object 
of this $i$ magnitude \markcite{sdss153}({Pier} {et~al.} 2002).
Pending the final definition of the SDSS photometric system,
the magnitudes are preliminary and are denoted by $u^*g^*r^*i^*z^*$ 
(see \S\,4.5 of \markcite{sdss85}{Stoughton} {et~al.} 2002).
All SDSS magnitudes are asinh magnitudes \markcite{sdss26}({Lupton}, {Gunn}, \& {Szalay} 1999), with zero-flux
magnitudes 24.63, 25.11, 24.80, 24.36 and 22.83 respectively.
}
\end{deluxetable}

\end{document}